\documentstyle[12pt]{article}
\begin{document}
\title{Elementary derivation of a recently proposed integral representation
for permanents}
\author
{Kacper Zalewski\thanks{Supported in part by the KBN grant 2P302-076-07}\\
Jagiellonian University\\ and\\ Institute of Nuclear Physcis, Krak\'ow,
Poland}
\maketitle

\abstract
{A recently proposed integral representation for permanents is rederived using
only elementary combinatorics. For this proof the assumption that the matrix,
for which the permanent is calculated, has an inverse is not necessary.}
\vspace{2cm}

The permanent of an $N\times N$ matrix $A$ is the number

\begin{equation}
P_N = \sum_Q \prod_{i=1}^N A_{i\;Qi},
\end{equation}
where $Q$ is a permutation of the numbers $i=1,2,\ldots,N$ and the summation
extends over all the $N!$ permutations. This formula is similar to the formula
definig the determinant of matrix $A$, but the minus signs in front of the
terms where the permutations $Q$ are odd are missing. In practice this makes
the handling of permanents more cumbersome than the handling of determinants.
The numerical evaluation of permanents, even of moderate size, is usually quite
hard. This is unfortunate, because they occur in some physical problems e.g. in
the description of the Bose-Einstein correlations in high energy multiple
particle production processes. The general experience is that even big
computers get in trouble with the exact formula soon after $N$ exceeds ten,
which is to soon for the applications.

A new approach to the evaluation of permanents has been recently proposed by
Wosiek \cite{JWO}, who pointed out that, when the inverse of matrix $A$ exists,
permanent (1) is equal to the integral

\begin{equation}
P_N = 2^{-N}\int\;\prod_{i=i}^N \frac{dxdy}{2\pi}\;e^{-\frac{x_i^2 + y_i^2}{2}}
\;\left[\left(\sum_{k=1}^N e_{ik}x_k\right)^2 + \left(\sum_{k=1}^N
e_{ik}y_k\right)^2\right],
\end{equation}
where matrix $e$, which is defined below, is constructed from the eigenvectors
and eigenvalues of matrix $A$. Here and in the following it is assumed that
matrix $A$ is real and symmetric. Since there are many methods of approximating
complicated integrals -- e.g. Monte Carlo methods and steepest descent methods
-- it is plausible that it is easier to get a good approximation to $P_N$ when
starting from formula (2) then when starting from formula (1). The examples
given in ref \cite{JWO} are very encouraging. In ref. \cite{JWO} formula (2)
was derived using methods inspired by quantum field theory. In the present note
we give an alternative, elementary proof. For this proof the assumption that
matrix $A$ has an inverse is no more necessary.

Let us note first that, since matrix $A$ is real and symmetric, it can be
diagonalized by an orthogonal transformation

\begin{equation}
O^T A\,O = \Lambda,
\end{equation}
where $\Lambda$ is a diagonal matrix with diagonal elements
$\lambda_1,\ldots,\lambda_N$. There may be many orthogonal matrices satisfying
equation (3). In such cases any of them can be chosen. Multiplying equality (3)
by matrix $O$ from the left and by matrix $O^T$ from the right, defining
\cite{JWO}

\begin{equation}
e_{ij} = O_{ij} \sqrt{\lambda_j}
\end{equation}
and using the orgthogonality property $OO^T = O^TO = 1$ we find

\begin{equation}
A_{ij} = \sum_{k=1}^N e_{ik} e_{jk}.
\end{equation}
Note further that the integral in formla (2) is a linear combination of
products of integrals of the type

\begin{equation}
\langle x^{2n} \rangle \equiv \frac{1}{\sqrt{2\pi}} \int_{-\infty}^{\infty}dx\;
e^{-\frac{1}{2}x^2}\;x^{2n} = (2n - 1)!!\;\;\mbox{ for }\;\; n \geq 1.
\end{equation}
In particular $\langle x^2 \rangle = 1$. Of course $\langle 1 \rangle = 1$ and
for any integer nonnegative $n$ --- $\langle x^{2n+1} \rangle = 0$,

We shall also use the fact that to each term in sum (1) corresponds a diagram
consisting of $N$ labelled vertices connected by $N$ lines, the $i$-th line
connecting the vertices $i$ and $Qi$. Note that a line can connect a vertex to
itself. Since each permutation can be decomposed into cycles, each diagram
consists of closed loops. For loops containing more than two vertices the two
orientations of the loop are distinguishable. Thus e.g. for $N=3$ the term
$A_{11}A_{22}A_{33}$ corresponds to three isolated vertices (one-vertex loops),
the terms $A_{11}A_{23}A_{32},\;A_{22}A_{13}A_{31}$ and $A_{33}A_{12}A_{21}$ to
the three diagrams consisting each of an isolated vertex and a two-vertex loop
and the terms $A_{12}A_{23}A_{31}$ and $A_{13}A_{32}A_{21}$ to the two
three-vertex loops with different orientations. Ascribing to each diagram the
corresponding product of matrix elements we find that the permanent is the sum
of the contributions corresponding to all the diagrams. For a symmetric matrix
$A$ the contributions corresponding to loops differing only by their
orientation are equal. Thus, an alternative approach is to ignore the loop
orientations and to ascribe an additional factor of two to each loop containing
more than two vertices. We shall need somewhat more complicated diagrams. Let
us label also the lines and ascribe to line labelled $k$ and connecting
vertices $i$ and $Qi$ the number $e_{ik}e_{Qi\,k}$. For such labelled diagrams
it is still true that the permanent is the sum of contributions corrresponding
to all the diagrams.

In order to evaluate the integral (2) let us consider first the terms in the
expansion of the integrand, where only the variables $x_1,\ldots,x_N$ occur and
where each of them occurs exactly twice. In this case each of the integrals (6)
equals one and we obtain the contribution to $P_N$

\begin{equation}
\Pi_N^0 = 2^{-N} \sum_Q \prod_{i=1}^N{}'\sum_{k=1}^N e_{ik}e_{Qi\,k}
2^{N-L(Q)},
\end{equation}
where $L(Q)$ is the number of cycles in permutation Q, or equivalently the
number of loops in the corresponding diagram, the prime over the product sign
means that only the terms where each value of the index $k$ occurs exactly
twice should be kept. In terms of labelled diagrams, only the diagrams where no
two lines have the same label are included and the contribution corresponding
to each diagram is reduced by the factor $2^{-L(Q)}$. This factor arises as
follows. Consider a loop with $M$ vertices. Each vertex corresponds to one
factor from the product in the integrand of formula (2). For $M=1$ and $M=2$
there are respectively one and two ways of choosing the necessary
$x$-variables. For $M>2$ at each vertex there are two ways of choosing the
necessary $x$-variable, thus the factor is $2^M$. This, however should be
divided between the two loops differing by their orientation. Thus for each
loop the factor is $2^{M-1}$. It is seen that this formula works also for
$M=1,2$. Evaluating the product of the factors corresponding to all the loops
in the diagram we find the overall factor $2^{N - L(Q)}$ as given in formula
(7).

As mentioned, the diagrams corresponding to formula (7) differ from the
diagrams corresponding to formula (1) in two ways. Firstly, the diagrams where
groups of lines have equal labels are missing. Secondly, there is an additional
factor of $2^{-L(Q)}$ for each diagram. In order to remove the first difference
let us start with a diagram, where a group of p lines has different labels and
look at the effect of ascribing the same label $l$ to these $p$ lines. It is
assumed that none of the other lines has either the label $l$ or the label of
any of the original $p$ lines. We will show that as a result the number of
diagrams increases by a factor $(2p-1)!!$. Consider the $2p$ vertices at the
ends of the $p$ lines. Of course some labelled vertices may occur in this list
twice. All the different connections of these $2p$ vertices by the $p$ lines
can be counted as follows. Consider the $(2p)!$ permutations of the $2p$
vertices and in each permutation connect with a line labelled $l$ the first
vertex with the second, the third with the fourth etc. This gives all the
connections, but with much multiple counting. The $p!$ permutations of the
lines among themselves and the $2^p$ exchanges of the ends of the lines do not
change the diagram. Thus finally there are

\begin{equation}
\frac{(2p)!}{2^p p!} = (2p - 1)!!
\end{equation}
new diagrams for the single original one, which proves our statement. On the
other hand, in the integrand this change introduces a factor $\langle x^{2p}
\rangle = (2p - 1)!!$, which is just enough to give weight one to each of the
new diagrams. Aplying repeatedly this argument it is seen that summing all the
terms, where the expansion of the integrand contains the variables
$x_1,\ldots,x_N$ only, one obtains

\begin{equation}
\Pi_N = \sum_Q \prod_{i=1}^N \sum_{k=1}^N e_{ik} e_{Qi\,k} 2^{-L(Q)},
\end{equation}
where the restriction that different line should carry different labels has
been removed. The factors $2^{-L(Q)}$ is justified just like in the case when
all the lines have different labels.

Let us consider now the contributions from the terms containing the $y$
variables. Including such terms it is easily seen that the diagrams are as
before, but that a loop can originate either only from the $x$ integrations, or
only from the $y$ integrations. Thus, the result of including the $y$ variables
is that each loop should be counted twice, or equivalently that it gets an
additional factor of two in its weight. For a complete diagram this yields the
additional factor $2^{L(Q)}$, which exactly cancels the unwanted factors in
formula (9). This completes the proof of formula (2).


\begin{thebibliography}{999}
\bibitem{JWO}J. Wosiek, {\it A simple formula for Bose-Einstein corrections,}
Jagel\-lonian University preprint TPJU 1/97 and Phys. Letters {\bf B} in print.

\end{thebibliography}
\end{document}